\documentclass[preprint]{aastex}
\usepackage{emulateapj5}
\usepackage{amsmath}
\usepackage{natbib}

\begin{document}

\received{}
\accepted{}

\slugcomment{Accepted by the Astronomical Journal}

\title{Assessing the Formation Scenarios for the Double Nucleus of M31 Using
Two-Dimensional Image Decomposition\altaffilmark{1}}

\author {Chien Y. Peng\altaffilmark{2}}

\altaffiltext{1} {Based on observations with the NASA/ESA {\it Hubble Space
Telescope}, obtained at the Space Telescope Science Institute, which is
operated by AURA, Inc., under NASA contract NAS 5-26555.}

\altaffiltext{2}{Steward Observatory, University of Arizona, 933 N. Cherry
Ave., Tucson, AZ 85721.  cyp@as.arizona.edu}

\begin {abstract}

The double nucleus geometry of M31 is currently best explained by the
eccentric disk hypothesis of Tremaine, but whether the eccentric disk resulted
from the tidal disruption of an inbounding star cluster by a nuclear black
hole, or by an $m=1$ perturbation of a native nuclear disk, remains debatable.
I perform detailed 2-D decomposition of the M31 double nucleus in the {\it
Hubble Space Telescope} $V$-band to study the bulge structure and to address
competing formation scenarios of the eccentric disk.  I deblend the double
nucleus (P1 and P2) and the bulge simultaneously using five S\'ersic and one
Nuker components.  P1 and P2 appear to be embedded inside an intermediate
component ($r_e=3\farcs2$) that is nearly spherical ($q=0.97\pm0.02$), while
the main galaxy bulge is more elliptical ($q=0.81\pm0.01$).  The spherical
bulge mass ($2.8\times10^7 M_\odot$), being coincident with the supermassive
black hole mass ($3\times10^7 M_\odot$), conjoined with a shallow bulge cusp,
are consistent with the scenario that the bulge was scoured by spiraling
binary supermassive black holes.  In the 2-D decomposition, the bulge is
consistent with being centered near the UV peak of P2, but the exact position
is difficult to pinpoint because of dust in the bulge.  P1 and P2 are
comparable in mass.  Within a radius $r=1\arcsec$ of P2, the relative mass
fraction of the nuclear components is $M_\bullet:M_{bulge}:P1: P2 =
4.3:1.2:1:0.7$, assuming the luminous components have a common mass-to-light
ratio of 5.7.  The eccentric disk as a whole (P1+P2) is massive,
$M\approx2.1\times10^7M_\odot$, comparable to the black hole and the local
bulge mass.  As such, the eccentric disk could not have been formed entirely
out of stars that were stripped from an inbounding star cluster.  Hence, the
more favored scenario is that of a disk formed in situ by an $m=1$
perturbation, caused possibly by the passing of a giant molecular cloud, or
the passing/accretion of a small globular cluster.

\end {abstract}


\section {INTRODUCTION}

M31 is the nearest galaxy with a normal bulge where there is good evidence for
the existence of a supermassive black hole (SBH) of $M_\bullet\approx
3\times10^7M_\odot$ (e.g.  Kormendy \& Bender 1999; Kormendy 1987; Dressler
\& Richstone 1988; Richstone, Bower, \& Dressler 1990).  Early Stratosphere II
balloon observations (Light et al.  1974) saw that M31 appeared to have a
dense nucleus embedded on top a galaxy bulge.  Nieto et al. (1986) showed that
the nucleus was off centered slightly from the bulge and asymmetric, which was
hinted at in Light et al. (1974).  It was not until {\it HST (Hubble Space
Telescope)}\ WFPC (Wide Field and Planetary Camera) and FOC (Faint Object
Camera) observations, that the nucleus was resolved into double components
(Lauer et al. 1993 [L93]; King, Stanford, \& Crane 1995 [K95]; Lauer et al.
1998 [L98]).

Figure 1 shows a grey scale $V$-band $(F555W)$ image of the nucleus from {\it
HST}\ WFPC2 in $V$, which was deconvolved and kindly provided by T. Lauer (see
L98).  The brighter of the double peak in the $V$-filter is designated as P1,
and it is separated from the fainter P2 by $0\farcs49$, or 1.8 pc (L93).  In
the far-UV, however, K95 discover that P2 is actually more luminous than P1.
There is a striking UV peak (upturn) on top of P2 which stands apart in color
from both P2 and P1.  In literature, P2 is synonymous with the UV peak, but
here I make the distinction because the peak is not centered on P2, as shown
later.  In the infrared, the contrast between the two nucleus is not quite as
strong as in the optical, and the peak seen in UV-light all but disappears,
either because it is intrinsically faint or is smeared away by the PSF
(Corbin, O'Neil, \& Rieke 2001).  Dust is ruled out as being the cause for the
appearance and asymmetry of the double nucleus based on color images of the
optical, far-UV, and near-infrared H-K (Corbin, O'Neil, \& Rieke 2001; Mould
et al. 1989), as well as a two-dimensional decomposition by K95, who impose
the assumption that both P1 and P2 components are smooth.

Despite the prominence of P1 in optical images, L93 and L98 show that it is
the UV peak which corresponds more closely to the center of the galaxy bulge,
using isophotal centroids.  They also suggest it is the location of the SBH.
From rotation curves, Kormendy \& Bender (1999 [KB99]), Bacon et al. (2001
[BA01]), and Statler et al. (1999 [S99]) find that the UV peak of P2 also lies
close to the dynamical center.  The most precise measurements put the
zero-velocity center at $0\farcs051\pm0\farcs014$ (KB99) and $0\farcs031$
(BA01) from the UV peak of P2, in the direction of P1.  However, the peak of
the velocity dispersion is $0\farcs2$ from P2, {\it away} from P1 (KB99,
BA01).  KB99 locate the SBH near the center of the UV peak.

K95 speculate that the strong UV peak may be due to a non-thermal, low-level,
active galactic nucleus (AGN), coincident with X-ray and radio emissions, but
acknowledge it is unclear whether it is truly resolved in pre-COSTAR {\it
HST}\ images.  Moreover, it has the energy output of only a single post
asymptotic giant star (K95).  L98 find that the UV peak is resolved by {\it
HST}\ WFPC2, leading them to hypothesize that the bluish color may not be an
AGN, but instead is caused by a population of late B to early A-type stars in
a cluster, possibly in a region with high metallicity abundance (e.g.
Burstein et al.  1988, and K95).  Forming such a cluster may be difficult,
however, because of the tidal influence of a nearby $3\times10^7M_\odot$ black
hole.  The AGN speculation about the UV peak has not been clarified despite
Chandra observations.  Garcia et al.  (2000) discover X-ray emission in
proximity to P2 with an unusual spectrum indicating possible relation to an
active nucleus.  The error circle, however, is 1\arcsec (twice the separation
between P1 and P2), and is not alone -- there are a number of other X-ray
sources farther away around the bulge.  If the UV peak is in fact AGN related,
it is curious that it does not have a counterpart in the near-IR where the
emissions are often seen more clearly than in the optical due to a typical
rise in the AGN spectrum and diminished extinction (Quillen et al.  2001).

Several empirical evidences indicate that the nucleus and the bulge have a
different formation history.  Within $0\farcs4\times0\farcs4$ of P1, L98 show
that the color is redder in $V-I$ ($=1.41$) than the surrounding bulge
($V-I=1.34$).  But in $U-V$, P1 (=2.29) and its anti-P2 direction ($U-V$=2.18)
appear to be bluer than the bulge ($U-V$=2.39).  This is in contrast to BA01
who find both P1 and anti-P2 to also be redder.  Nonetheless, both their
findings indicate that the nucleus has a stellar population, metallicity, and
perhaps origins, different from that of stars in the bulge.  Sil'chenko,
Burenkov, \& Vlasyuk (1998 [SBV98]) show that the nucleus is more metal rich
than the surrounding.  In the two regions, the spectral indices differ by,
$\Delta {\rm Mg}_{\rm b}=0.86\pm0.1$ and $\Delta \left<{\rm Fe}\right>=0.53\pm
0.08$.  KB99 confirm that P1 and P2 have a similar stellar population, have
higher metal line strengths, and are more similar to each other than to the
bulge or to any globular cluster.  SBV98 also use $[{\rm H}\beta/\left<{\rm
Fe}\right>]$ to show that stars in the nucleus are younger by roughly
$\lesssim8$ Gyrs.

The compactness of the P1 and P2 peaks is particularly interesting and
challenging to explain from the standpoint of galaxy formation and dynamical
evolution.  KB99 determine the combined mass of the double nucleus to be
$M_{nuc}\simeq 3.5\times10^7 M_\odot$, comparable to the mass of the SBH.
That such a configuration exists at all with such close separation (1.7 pc) is
surprising, and a natural interpretation is that P1 is either a merger remnant
or a captured star cluster.  Emsellem \& Combes (1997) model this scenario
with N-body simulations in a SBH potential and find that, although they can
reproduce the geometry, the lifetime is on the order of 0.5 Myr.  If the
cluster is massive enough to survive disruption, in most cases, dynamical
friction would cause P1 to spiral in and coalesce with P2 within several
dynamical times, unless somehow P1 co-rotates with the bulge (K95).  Although
the short life span of an accreting scenario is not in itself conclusive
proof, the discovery of other similar such systems (e.g. NGC 4486B, Lauer et
al.  1996) favors a scenario which has a much longer lifespan.  Moreover, S99
find that the sinking star cluster of Emsellem \& Combes (1997) model does not
reproduce either the rotation curve or the dispersion profile.

Tremaine (1995) proposes an eccentric disk model to explain the nuclear
configuration which currently stands as the favorite model.  The thick
eccentric disk, nearly Keplerian, is composed of stars in ring orbits around a
black hole, and P1 is caused by a bottleneck of stars slowing down at the
turnaround radius (apocenter).  This model also predicts that the velocity
center is displaced by $\sim 0\farcs2$ from P2 in the direction of P1.
Moreover, it predicts that the stars in P1 and P2 should have a similar
stellar population because they belong to the same system, which might not be
the case if P1 is a merger remnant.  With spectroscopic data, KB99 show that
the shift of the rotation center is in the right sense of the prediction, even
if the amount is somewhat lower than prediction.  Moreover, P1 and P2 have
more similar stellar populations to each other than to the bulge in accord
with the model prediction.  However, the missing element in the Tremaine model
is an explanation for how the ring orbits maintain alignment under precession:
the original model has no self gravity.  Since then, several workers have
proposed enhancements.  Among them, Salow \& Statler (2001) propose a
semi-analytic eccentric disk with self-gravity around a black hole, where the
disk is made up of a superposing set of Keplerian orbits dispersed in
eccentricity and orientation according to a certain distribution function.
It predicts that the line-of-sight velocity distributions of the disk near
the black hole should have a distinctive double peak, which provides a further
observational constraint to test the eccentric disk model.  Using a different
formalism, with integrable models, Jalai \& Rafiee (2001) show that the double
nucleus geometry can be sustained with four general types of regular orbits in
a St\"ackel potential, even in the absence of a SBH.  However, the requirement
that the nuclei be cuspy, with surface density $\Sigma\propto r^{-2}$ is a
stringent requirement, and is not seen in M31.  Furthermore, it is not clear
why the nucleus would be asymmetric.

Bekki (2000 [BE00]), BA01, and Sambhus \& Sridhar (2001 [SS01]), propose
scenarios for how such an eccentric disk might be formed and sustained under
precession with self-gravity.  In BE00 N-body simulation, an inward bound
$10^6M_\odot$ star cluster is completely disrupted by the tidal shear of a
$10^7M_\odot$ black hole, consequently forming a thick, eccentric, stellar
system.  The alternative scenario by BA01 involves a circular disk, already
present in the bulge, that gets excited into an eccentric disk by a natural
$m=1$ perturbation.  The excitation may either be caused by a passing giant
molecular cloud or globular cluster, which can induce the lopsidedness in the
disk (seen in Fig.~1) that persists for $7\times10^7$ years in their
simulations.  However, for the non-axisymmetric waves to develop, the disk
needs to be thin.  They furthered considered a model in which a black hole was
shifted from the center of the potential, and another where its velocity was
perturbed.  Too many particles escaped in the simulations for them to be
viable.  The model proposed by Sambhus \& Sridhar (2001) extends Tremaine's
(1995) eccentric disk to simulate a larger disk mass ($2\times10^7 M_\odot$)
which involves a shredding of $10^5M_\odot$ globular cluster near the vicinity
of the SBH.  Subsequently, the cluster stars merge into a pre-existing disk of
few $10^7 M_\odot$, orbiting the central black hole on both prograde and
retrograde, quasi-periodic, loop orbits.  The $m=1$ instability which causes
the large eccentricity is induced by resonant response to the counter-rotating
orbits.  In this model, lopsidedness of the eccentric disk geometry is created
in response to the presence of retrograde orbits.  All three models roughly
reproduce the double nucleus geometry and the dynamics, such as the offset in
the velocity dispersion profile from P2 by $\approx 0\farcs2$, to varying
degrees of accuracy.  However, one important difference between BE00 and BA01
+ SS01 scenarios is that the BE00 model requires the stellar cluster be much
less massive than ($\lesssim 10\%$) the black hole so that it can be
disrupted.  By contrast, the BA01 and SS01 models are constructed to have a
much larger disk, about $20\%-40\%$ of the total central mass concentration.
These two scenarios can be directly tested if a mass estimate of the disk can
be robustly measured, and the uncertainty can be quantified.  The ambiguity in
the bulge decomposition is reflected in the available photometry of P1, which
can significantly differ in brightness from different studies.  Furthermore,
thus far it is not clear how much of P2 is part of the bulge or the eccentric
disk.

In addition to the eccentric disk, the bulge structure of M31 and how it fits
into the developing picture of galaxy formation are interesting on their own
merits.  However, their studies have been complicated by the double nucleus.
The steepness of galaxy nuclei and correlations with other structural
parameters (e.g. Faber et al.  1997) reflect the manner by which bulges are
formed.  Numerical simulations show that black hole mergers can flatten galaxy
cores by ejecting stars from the center (e.g. Ebisuzaki, Makino, \& Okumura
1991; Nakano \& Makino 1999; and Milosavljevi\'c \& Merritt 2001).  This
scenario appears promising for explaining the correlation found between large
bulges and low central surface brightness.  In this paper, I study the
detailed properties of the M31 bulge by decomposing optical images to provide
new structural parameters.  With detailed decomposition one can address the
following issues:  Are there subtle structures in the bulge that are not
obvious in full light?  What are the relative contributions of the bulge, P1,
and P2 components?  What are their shapes?  What is the bulge profile and how
sharply is it peaked?  Finally, I discuss what the new photometry of P1 and P2
reveal about the two competing scenarios that explain the formation of the
eccentric disk.  Although some of these questions can be addressed from other
data in previous studies, this new analysis provides a unique look at the
double nucleus based on a more flexible range of assumptions than foregoing
studies.

In the sections to follow, Section 2 discusses the data, Section 3 briefly
describes the analysis algorithm used to deblend the bulge.  Section 4
discusses the decomposition, followed by the environment of the bulge in
Section 5.  Section 6 compares the eccentric disk formation models.
Conclusions follow in Section 7.

Throughout the discussion I assume that the distance to M31 is $D=770$ kpc,
following KB99.  I also assume that the Galactic extinction is
$A_{\mbox{v}}=0.24$ (Burstein \& Heiles 1984), which is similar to
$A_{\mbox{v}}=0.21$ determined by Schlegel et al. (1998). Kormendy (1988) and
KB99 dynamical models also suggest that $M/L \approx 5.7$ for the bulge stars.
This is similar to star formation models of Bell \& de Jong (2001), from which
one can derive $M/L\approx 6$ based on the $V-I$ color of the bulge.

\section {Data}

I obtain $V$-band ($\approx F555W$) data (GO 5236, Westphal) from the {\it
HST}\ archive, as well as the deconvolved version from T. Lauer.  The two
sets of images are used for different purposes.  The deconvolved version has
been trimmed to a smaller field of view (FOV) of
$11\farcs8\arcsec\times11\farcs8$.  While ideal for studying the nucleus,
there is not enough realty for measuring the bulge profile and its shape.
Therefore I also create a mosaic image with a FOV of
$150\arcsec\times150\arcsec$, with a missing echelon in the Planetary Camera
(PC) quadrant.  The net exposure time is 300s, which has sufficiently high
signal-to-noise (S/N) to measure the bulge profile.  In addition to the
deconvolved and mosaic images, I create a dithered image of the PC chip (FOV =
$35\arcsec\times35\arcsec$) by combining four exposures of 300 second images.
This will be used later for comparisons with the deconvolved image.

\section {Technique}

\subsection {Algorithm}

I use a general galaxy fitting program called GALFIT to do the 2-D
decomposition.  Detailed information on the software and how it is implemented
are found in a companion paper by Peng, Ho, Impey, \& Rix (2002), but here I
describe it briefly.  One of the design capabilities of GALFIT is to
accurately decompose nearby galaxies that are highly resolved to study them
closely and uncover or extract galaxy sub-structures, such as nuclear disks,
bars, and double nuclei.  To be highly flexible, GALFIT allows simultaneous
fitting of standard function types such as S\'ersic (1968), Gaussian,
exponential, and Nuker.  The program has the option to either convolve the
models with the PSF to simulate the seeing, or not convolve if the image has
already been deconvolved.  The number of components to fit is not limited a
priori.  GALFIT minimizes $\chi^2$ residuals using a down-hill
gradient/parabolic expansion method called Levenberg-Marquardt (Press et al.
1997) by iteratively creating model images, convolving them with the PSF, and
subtracting them from the data.  Even though the gradient method is not as
``smart'' as alternative Simulated Annealing or Metropolis algorithms, it has
the virtue of being fast.  As an example, fitting 6 components with 41 free
parameters over the entire $400\times400$ pixel image, while doing
convolution, takes roughly one minute per iteration on a Pentium 450 MHz
computer, and converges in $30-50$ iterations.  Speed is desirable because the
$\chi^2$ topology of the fit is complex; to adequately explore it and find an
optimal fit thus requires testing various model combinations and initial
parameters.

The merit function to minimize is the $\chi^2$, or in normalized form,
$\chi^2_\nu = \chi^2/N_{\rm dof}$, defined as:

\begin{equation}
\chi^2_\nu = {\frac{1} {N_{\rm dof}}}\sum_{x=1}^{nx}\sum_{y=1}^{ny} \frac{{\left(\mbox{flux}_{x,y} - \mbox{model}_{x,y}\right)^2}} {\sigma_{x,y}^2}.
\end{equation}

\noindent where $\sigma_{x,y}$ is the uncertainty, or weight, at each pixel;
$N_{\rm dof}$ is the degree of freedom; $nx$ and $ny$ are the dimensions of
the image in $x$ and $y$ direction.  ${\mbox{Model}_{x,y}}$ is the sum of all
the model components fitted.  

Traditionally, 2-D models use purely elliptical azimuthal shapes, but most
galaxies are somewhat disky or boxy depending on whether they are more
rotationally or pressure supported, respectively.  GALFIT can fit them by
replacing the ellipses with a shape of,

\begin {equation}
r = \left(\left|x\right|^{c+2}+\left| \frac{y} {q}\right|^{c+2}\right)^{\frac{1}{c+2}},
\end{equation}

\noindent which can be rotated to any position angle (PA).  The component
shape is boxy if $c>0$, disky if $c<0$, and pure ellipse when $c=0$.  This
azimuthal function was motivated originally by Athanassoula et al.  (1990) to
characterize the shapes of galaxy bars.  The parameter $q$ is the axis ratio
of each component.

In the radial direction, the 1-D S\'ersic light profile is defined as,

\begin {equation}
\Sigma(r)=\Sigma_e \mbox{ exp}^{-\kappa \left[({\frac{r}{r_e}})^{1/n} - 1\right]}
\end {equation}

\noindent where $r_e$ is the effective radius of the galaxy, $\Sigma_e$ is the
surface brightness at $r_e$, $n$ is the power-law index, and $\kappa$ is
coupled to $n$ such that half of the total flux is within $r_e$.  The original
de~Vaucouleurs (1948) profile is the special case with $n=4$ and
$\kappa=7.67$.  The usefulness of the S\'ersic function is that it can fit a
continuum of profiles ranging from a Gaussian, to exponential, to a
de~Vaucouleurs, by smoothly varying the exponent: $n=0.5$ for a Gaussian,
$n=1$ for an exponential disk, and $n=4$ for de~Vaucouleurs.  In GALFIT, the
flux parameter that is fitted is the total magnitude, instead of $\Sigma_e$.
The translation between the total magnitude of a component and $\Sigma_e$ is
given in Peng et al. (2002).

Bulges that are modeled with the S\'ersic profile have flat cores, but most
galaxies have profile rise that extend into the resolution limit (e.g. Lauer
et al. 1995, Carollo et al. 1997, Rest et al.  2001, and Ravindranath et al
2001).  A function which can describe many double power-law and core type
galaxy bulges in 1-D was introduced by Lauer et al.  (1995), known as the
Nuker law:

\begin{equation}
I(r) = {I_b \ 2^{ \frac{\beta - \gamma}{\alpha}}
\left({\frac{r}{r_b}}\right)^{-\gamma}\left[{1+\left({\frac{r}
{r_b}}\right)^{\alpha}} \right] ^{\frac{\gamma-\beta}{\alpha}}}
\end{equation}

\noindent There are five adjustable profile shape parameters: $I_b, r_b, 
\alpha, \beta,$ and $\gamma$.  Taken to the limits of large and small radii,
the parameter $\gamma$ is the slope of the inner power law, and $\beta$ is the
slope of the outer power law.  The break radius $r_b$ is the location where
the profile changes slope, $I_b$ is the surface brightness at $r_b$, and
$\alpha$ describes how sharply the two power laws connect.  GALFIT fits the
surface brightness magnitude $\mu_b$ instead of the intensity $I_b$.  I
generalize this profile to 2-D, hence there are 5 additional parameters:
$x_{cent}, y_{cent}, q$, PA, and $c$.

To account for the PSF seeing when fitting models to the dithered image,
GALFIT convolves a doubly oversampled PSF, created using Tiny Tim (Krist \&
Hook 1997), with the models.  The Tiny Tim software can reproduce the core of
WFPC2 PSFs well, even if the diffraction spikes are harder to reproduce.  For
convolution purposes this should suffice.  Later in the paper, I will compare
these results with fits to the deconvolved image of L98.

\subsection {Noise Image and Parameter Uncertainty Estimates}

Locally, the bigger of the two contributors to the root-mean-square (RMS)
fluctuation of the data image is the surface brightness fluctuations (SBF) of
star clusters resolved by the {\it HST}, rather than the Poisson noise.
Unlike Poisson noise where the noise-to-signal ratio decreases as $\sqrt{\rm
time}$, the ratio of the SBF noise to the galaxy background asymptotically
approaches a constant value.  Thus, in this regime, the derived uncertainties
in the fitting parameters remains constant with increasing exposure time.
Moreover, while the Poisson noise correlates on scale of $\lesssim 1$ pixel in
the dithered image, the SBF noise correlates on scales of a few pixels.  To
get realistic error bars from model fits thus requires having a reasonable
estimate of the pixel weights, $\sigma_{x,y}$, that account for the SBF in a
steeply rising galaxy bulge.

A Poisson map of the image bears an imprint of the SBF variance because the
same star clusters that produce the Poisson signal are responsible for the
SBF.  I assume that the SBF RMS can be approximated by a scaling of the
Poisson noise.  To do so, I compute the ratio of the total RMS to the Poisson
noise in $30\times30$ pixel boxes sampled randomly around the bulge in the
dithered image.  Then, multiplying this factor (1.9) to the Poisson map of the
image, I obtain an approximate SBF noise image to use as pixel weights.  The
exact scale factor does not affect the estimated uncertainties because they
are renormalized by $\sqrt{\chi^2_\nu}$ (see below).  This scale factor
appears reasonable, and for visually good fits, $\chi^2_\nu$ comes out close
to unity.

It is worth noting that the intrinsic galaxy SBF amplitude is higher than in
the observed image because of PSF convolution.  Moreover, because the SBFs are
spatially resolved, PSF smearing correlates the SBF noise so that the pixels
are not statistically independent.  However, in scaling the pixel weights as
described, the PSF correlation of the SBF is implicitly taken into account in
the $\chi^2_\nu$ statistics.  To first order, the PSF smearing effectively
reduces the $N_{\rm dof}$ in the observed image by a factor $C$.  This factor
is $C\sim{\rm var}_{int} / {\rm var}_{obs}$, where ${\rm var}_{int}$ is the
intrinsic variance (i.e.  without PSF filtering) and ${\rm var}_{obs}$ the
observed variance (i.e. PSF-filtered).  Furthermore, the effective pixel
weights $\sigma_{x,y}^2$ have the same factor $C$, which cancels out of
Equation [1] in calculating $\chi^2_\nu$.

Once GALFIT converges on a solution, it estimates the fitting uncertainties by
using the covariance matrix of the parameters.  For a more detailed
discussion, see Peng et al. (2002), but here I give a brief description.
Since $\chi^2$ is a function of the fitting parameters, the surface of
constant $\Delta\chi^2$ centered around the best fit $\chi^2_{min}$ can be
roughly approximated as an {\it n}-dimensional ellipsoid, where {\it n} is the
number of free parameters in the fit.  Defined as the local $\chi^2$ curvature
with respect to all pairs of parameter combinations $a_i$ and $a_j$, i.e.
$\mbox{d}^2\chi^2/(\mbox{d}a_i \mbox{ d}a_j)$, the covariance matrix of the
fit provides a handle on the correlation between parameters.  To estimate
parameter uncertainties accounting for correlation, I consider an ellipsoid
which encloses a region up to $\Delta\chi^2=1$, which is the 68\% confidence
region for a single degree of freedom.  The ellipsoid semi-major axes and
their lengths are calculated from the eigenvectors and eigenvalues of the
covariance matrix.  I determine the uncertainty $\sigma a_i$ for parameter
$a_i$ by projecting the vector sum of all the major axes vectors onto the unit
coordinate axis $\hat{a}_i$.  In cases when the fit does not produce a reduced
$\chi^2_\nu\approx1$, I also scale the uncertainties by $\sqrt{\chi^2_\nu}$ to
avoid under-estimating them.  Strictly speaking, while this method of
estimating uncertainties is fair, it may produce uncertainties that are equal
to or greater than the true uncertainties.  This over-estimate is probably not
a serious concern because, ultimately, the limitation is that we do not know
how well the local $\chi^2$ topology may be represented by an ellipse.
Furthermore, the parameter values may not be normally distributed around the
mean.

In the discussions to follow, I quote $\chi^2_\nu$ and parameter values based
on fits to dithered images rather than deconvolved images.  Although the image
of M31, with a very high S/N, has been deconvolved accurately, the process
introduces an additional correlation by the PSF, which is hard to interpret in
the presence of SBF.  In comparison, the noise property of the unrestored
image is easier to interpret.  While the signal in individual, unrestored,
images are correlated by the PSF from pixel to pixel, the Poisson noise is
{\it uncorrelated}.  Combining dithered images introduces at most a sub-pixel
correlation which is small compared to the SBF.  The correlation in the signal
is accounted for in the fitting process by convolving the models with the PSF.

\section {Decomposition}

My attempt at decomposing the double nucleus is not the first of its kind, but
is motivated by previous studies of L93, K95, and BA01, who perform 2-D image
decomposition using iterative techniques.  K95 fit the $U$-band image by first
subtracting a smooth model of P2 from the data.  Then, fitting isophotes to P1
and obtaining a smooth model, they use it to get a better estimate of P2.  The
process is then repeated.  BA01 decompose the WFPC2 $I$-band image by assuming
the bulge can be fitted by three Gaussians after masking out the central
$2\arcsec$ of the image.  In so doing, there is an implicit assumption that
the galaxy core flattens out.  The fit is then applied to $V$ and $U$-band
images after proper normalization.  L93 use a nonlinear least-squares routine
to deblend the nuclear region simultaneously with two components, assuming
that P2 is indistinguishable from the bulge.  Each component is concentric,
with otherwise arbitrary position, brightness, ellipticity, and position
angle.

My procedure is similar to L93 study, but with a more general set of
assumptions:  I relax the restriction that P1 and P2 are symmetric, and the
number of components needed to fit them.  The steepness of the bulge is a free
parameter.

\subsection {Fitting Procedures}

\subsubsection {General Outline}

M31, being an Sb-type galaxy, has a large galactic disk component that spans
$3.5^\circ$ in diameter, with a bulge that has an effective radius of
$4\farcm5$ (KB99).  Although the PC chip has twice the angular resolution to
resolve the double nucleus compared to the Wide Field chips, the combined
WFPC2 mosaic, with $\approx 3\farcm5$ FOV, is better suited for measuring the
large bulge.  To take advantage of the capabilities of both PC and WFPC2, I
perform fitting in two separate steps.

First I fit the bulge with a Nuker profile using the entire $V$-band mosaic
image.  To do so, I mask out the echelon-shaped region outside the PC, the
dust lanes in the mosaic image, as well as inner $3\arcsec$ of the nucleus.
Masking the nucleus does not seriously bias the bulge measurement because, in
the second step, I take the fitted parameters for the bulge, and optimize it
simultaneously along with other components for the nucleus in the PC chip.  In
this step the bulge position and inner slope parameter, $\gamma$, can vary
freely, but all other {\it bulge} parameters are fixed at their optimal
values.  All the other {\it nuclear} components can change freely without
constraints.  The results are discussed in \S~4.3, and summarized in Table 1.
As an independent check on the decomposition, I perform a similar fit by
replacing the Nuker bulge with a S\'ersic bulge, presented in \S~4.4.

I note that even though M31 has a large disk component, using 1-D
decomposition, KB99 show that the extrapolation of the disk into the center,
and within the WFPC2 FOV, is nearly constant at $\mu_V\approx20.4$
mag/arcsec$^2$.  Henceforth I adopt this value as a sky level in all fits.

\subsubsection {Determining the Number of Components}

I determine the number of components to fit iteratively by starting with the
bare-minimum assumption, then increase that number to see how the fit improves
$\chi^2$.  L93 has already found that two components failed to fit the M31
nucleus.  Therefore, my first trial starts with three, representing P1, P2,
and a bulge.  For a three component fit, the $\chi^2_\nu$ is 1.57, with
$N_{\rm dof} = 1.6\times10^5$.  In comparison, an excellent fit to the bulge
in the PC image, after masking out the double nucleus, produces a
significantly lower $\chi^2_\nu=1.25$.  The reason for the bad fit is that P1
is asymmetric (L93, K95, L98, and BA01), moreover, the UV peak is not centered
on P2 (BA01).  This can be gleaned from Figure 1 where the bright UV peak is
displaced to slightly south-east of P2, and P1 is not purely ellipsoidal.  To
model asymmetry, each nucleus requires at least another component nearby,
totaling five.  I use the S\'ersic profiles for most components, except for
the bulge.  For the bulge, I choose from either a S\'ersic or Nuker
component.

Fitting five components to just the PC image and not the mosaic, allowing {\it
all} parameter to optimize (40-42 parameters, depending on whether a S\'ersic
or Nuker function is used), one can obtain a good fit with a $\chi^2_\nu
=1.31$.  A closer examination reveals that the dominant component, i.e. the
bulge, in this trial has an axis ratio $q=0.90$ with uncertainty $<0.01$.  The
fit might initially be deemed acceptable except for the fact that the bulge is
noticeably elliptical in the mosaic image, with an axis ratio $q\approx0.80$,
and the inner nucleus is even more flattened.  Therefore, at an intermediate
radius, there must be a round component that dominates, and has significantly
different properties than the larger scale bulge.  I can derive meaningful
bulge parameters from the mosaic image as described previously in
\S~4.1.1, and apply it fixed just to the PC image alone, using five
components.  Here again, the component with $q\ge0.9$ reappears at a highly
significant level, at the expense of the UV peak component.  Inevitably, in a
realization using five components, the UV peak is poorly fitted because it
makes up only 0.1\%, compared to the round bulge component, which makes up as
much as 16\%, of the flux in the inner few arcseconds.  The round $q\ge0.9$
component can in principle be a stellar disk seen face on, but this is
unlikely.  It has a half light radius $2\farcs4$ to $3\arcsec$, roughly five
times the double nucleus separation.  Hence, from here on I group this
component with that of the bulge, but designate it as the ``spherical'' bulge
(fifth component in Table 1) to distinguish from the large scale bulge
component.  I will return to a more detailed discussion of the spherical
component in Section 4.6.

The best fit, therefore, requires six components, with $\chi^2_\nu=1.30$: two
for each nucleus, and two bulge-like components, to be discussed shortly.  To
fit P1 and P2, S\'ersic models do a fine job without having to use a more
flexible Nuker function.  Table 1 summarizes the properties of all the
components for two different trials.  In the first trial, I fit the dominant
galaxy bulge with a 2-D Nuker model, and in the second, with a S\'ersic model.
I discuss these two scenarios with more detail independently in \S~4.3 and
\S~4.4.

\subsection {Uniqueness and Meaning of the Components}

To fit the double nucleus accurately I ultimately resort to fitting the image
with six components, with 40-43 free parameters, which might seem
bewilderingly large and ill-constrained.  I emphasize however, that they may
not all be physically distinct components.  In particular two components are
used each for the bulge and P1 because they are not simple to parameterize.
Therefore, it may be more meaningful to interpret the sum of each pair as a
single component rather than each as separate entities.

With a high number of components, several issues can be raised regarding the
uniqueness and validity of decompositions (A. Quillen and T. Lauer, private
communication).  First, if the nuclei are related to an eccentric disk, its
light distribution may be irregular and may not be parameterizable.  So even
if the decomposition may look acceptable, the model will not be physically
meaningful.  Another objection is that the presence of the disk may perturb
the bulge enough kinematically to invalidate the assumption that the bulge is
smooth at the center.  Therefore, extrapolating the bulge profile into the
center based on outer regions may be unreliable.

I argue that there is evidence that the bulge profile is not disrupted much by
the process of the double nucleus formation:  Often, kinematics can reveal
subtle physical attributes (e.g.  kinematically decoupled core) that are not
obvious in the starlight.  Published rotation curves and velocity contours
show that the nucleus has a regular solid body rotation out to $0\farcs5$,
which becomes Keplerian beyond $1\arcsec$ (KB99, BA01).  The kinematic curves
are otherwise featureless apart from the velocity dispersion peak being offset
$0\farcs2$ from the UV peak away from P1.  Therefore, either the formation of
the disk has not been sufficiently violent to disarrange the nuclear dynamics,
or the perturbations have relaxed into the bulge.  Hence I believe the bulge
can be parameterized by smooth functions down to small physical scales.
Furthermore, I believe that the degree to which the double nucleus is
parameterizable and regular may yield insights into the eccentric disk and
formation model predictions on the morphology.  Being a complicated system,
irregularities in the nuclei would likely show up in the residuals because the
functions I fit have a high degree of symmetry.  Recognizing, however, that
inferences about the morphology of the double nucleus is intimately tied to
the bulge, I will check how assumptions about the bulge can affect the
decomposition.  As a prelude to discussions to come, I find that, throughout
various trials for the bulge, the sum of the parts for P1, P2, and the bulge,
result in fundamentally similar components, even if individual sub-components
may not be unique.  For these reasons I believe the decomposition to be valid
and robust.  But ultimately, as with K95 I let the smoothness and the pattern
of the residuals be an empirical guide to what is a reasonable solution.

\subsection {Nuker Bulge}

In their fit to a 1-D bulge profile, KB99 obtain the Nuker parameters:
$\mu_b=17.55\pm0.66$, $r_b=67\farcs7\pm62\farcs1$, $\alpha=1.08\pm0.59$,
$\beta=1.51\pm0.57$, $\gamma=0.25\pm0.25$, based on a fit to a region
$5\farcs5 \le r \le 300\arcsec$, excluding the double nucleus.  The large
uncertainties are due to parameter coupling in the Nuker profile. 
As initial parameters to my fit of the bulge, I use the values from KB99 but
allow them to vary.  

The best fit model I obtain, according to $\chi^2$, has a bulge that is
represented by a Nuker model, with an overall $\chi^2_\nu=1.30$.  In all, six
components are used, and they fit the entire bulge region well, as shown in
Figure 2.  Figure 2b shows the net residuals; Fig. 2c shows the model +
residual for P1, and similarly for P2 in Fig. 2d.  The images shown in Fig.  2
are fits to deconvolution restored image of L98, but the numbers presented in
Table 1, including $\chi^2_\nu$ are from the unrestored image for reasons in
\S~3.2.  In comparison to KB99, my fits to the mosaic and dithered PC images
produce parameters very similar to that obtained by KB99, to within their
uncertainties:  $\mu_b=17.76$, $r_b=66\farcs48$, $\alpha=1.10$, $\beta=1.99$,
$\gamma=0.17$.  My central cusp measurement, $\gamma=0.17$, is slightly
shallower than KB99, $\gamma=0.25$, in this more detailed fit to the double
nucleus.

In this model, the centroid of the bulge component is displaced from the UV
peak, by $(\Delta\mbox{RA}, \Delta\mbox{DEC})=(-0\farcs06, -0\farcs15)$, which
is a direction away from the dust patches in the bulge.  Such a large shift is
most likely due to disturbance by dust around the nucleus; forcing the bulge
to be centered on the UV peak increases the $\chi^2_\nu$ by 0.003.  In light
of the unknown effect by dust, this small difference may not be statistically
significant.

The residual error in Fig.~2b is 6\% on average at the location of the worst
fit, which is comparable to the SBF.  The parameters obtained using the
dithered and deconvolved images agree very well to within the uncertainties.
Figure 3 shows the residuals near the double nucleus divided by the RMS image
-- the calculation of which was discussed in \S~3.2.  The worst fit within the
1\arcsec\ circle is near P1 at ($-0\farcs22$, 0\farcs04), where one pixel
deviates by $-3.7\times$RMS.  Indeed, even though the fit is overall quite
good, there are differences between the model and the data.  In Figure 4, I
show the model image by summing P2 (without the UV peak) and the two
components of P1, together making up the eccentric disk.  The contour spacing
is 0.1 magnitude and the circle at (0\arcsec,0\arcsec) marks the location of
the UV peak.  Comparing Fig.~4 and Fig.~1 shows that although the asymmetry of
P1 is well modeled, P1 in the data is slightly rounder than the model, causing
an oversubtraction near ($-0\farcs22$, 0\farcs04).  Simultaneously fitting a
7th component to the image does reduce the ellipticity of P1 somewhat, and
reduces the $\chi^2_\nu$ by 0.02.  However, as the superficial improvement
does not affect the results below, I choose to present a simpler model using 6
components.

The major axis orientation of the eccentric disk shown in Fig.~4 is at
PA=$59.25^\circ$.  Assuming the disk is thin and circular at outer isophotes,
its inclination is $50\pm1^\circ$ to our line of sight.  The sum of the fluxes
from P1+P2 is $m_{\rm v} = 13.12\pm0.06$, or $M_{\rm v} = -11.55$ after
correcting for galactic extinction and distance.  The mass of P1+P2 is $M_{\rm
P1+P2}=2.1\times10^7M_\odot$.  In isolation, P1 has a brightness of $m_{\rm v}
= 13.65$ mag.

Figure 5 shows the radial surface brightness profile of the components
resulting from the decomposition (shifted to a common center), as well as
shape parameters from isophote fitting the entire WFPC2 mosaic {\it V}-band
image.  In that figure, I represent the individual components used in the fit
as dashed lines.  Because P1 is asymmetric, I only show the brighter of the
two sub-component, P1a.  The UV peak is not plotted.  The solid data points
shown with errorbars are the intrinsic profile of the bulge in the absence of
the double nucleus.  Exterior to $5\arcsec$, those points are measured from
isophote fits to the raw image, but interior, they have to be extrapolated
from the analytic model fits because of the double nucleus.  To do so, I
create an image of the bulge by summing the Nuker and spherical bulge
components, then fit it with isophotes.  For completeness, I also show the
shape parameters of the bulge out to 100\arcsec; the $cos\ 4\theta$ panel is a
measure of the diskiness ($> 0$) and boxiness ($<0$) of a given isophote.  The
bulge is only slightly boxy, evidenced in ${\rm cos}4\theta$ and from 2-D fits
($c=0.05$).  The presence of the spherical bulge component is manifest in the
dip in the ellipticity profile at $r=2\arcsec$.  Although the bulge profile in
general appears to be smooth with a gentle curvature farther out, with no
apparent breaks (see KB99), in the decomposition, there appears to be a break
at a radius of $1\farcs91$.  This kind of behavior is sometimes seen in large
bulges or spheroidal systems (e.g.  Lauer et al. 1995, Peng et al.  2002).  I
plot a 1-D Nuker function having the parameters shown in the figure to guide
the eye; it is not strictly a fit that minimizes the $\chi^2$.  I will discuss
this break in more detail in \S~5.

For the remainder of this sub-section, I test the extent to which P1 and P2
are affected by details of the bulge decomposition.  To put a limit on the
central cusp of the bulge I remove the UV peak model and instead fit the
bulge to that location and allow all the parameters to re-optimize.  The
profile of the UV peak differs significantly from that of the bulge, and the
$\chi^2_\nu$ increases by $\sim 0.01$ to 1.31, with $\gamma=0.19$.

In this model, the bulge component is centered on the UV peak.  However, P2
remains significantly off-centered from the UV peak by $(\Delta\mbox{RA},
\Delta\mbox{DEC})=(-0\farcs24, -0\farcs15)$.  I test to see if this is real or
is caused by fitting degeneracy by forcing P2 to the UV peak position.  I
refit both the deconvolved image of L98 and my own dithered image.  Both fits
produce residuals much worse than the asymmetric models.  In particular, in
the unrestored image the region in between P1 and P2 is significantly
over-subtracted.

To constrain the luminosities of P1 and P2, hence the eccentric disk itself, I
increase the nuclear slope parameter $\gamma$ from the best fit value.  If
$\gamma$ is forced up to 0.40, the fit significantly oversubtracts the bulge,
which can be safely ruled out.  This fit puts a lower limit on the brightness
of P1 to be 0.15 mag, and P2 to be 0.4 mag, fainter than the optimal fit.

In their decomposition, K95 find that P1 has a brightness of $m_{\rm v}=14.8$,
and in L93, they find 14.5.  My numbers are significantly brighter than both
by about 0.9 to 1.2 magnitude.  The differences may be in my assumptions about
the true shape of P1 and P2, as well as the relative contribution of the
bulge.  L93 assume that both P1 and P2 are individually symmetric and
concentric, with the bulge centered on P2.  Part of the remaining differences
might also be in the criteria of smoothness.  My criterion is formally more
strict than theirs.  L98 also find the blue UV peak to have $m_{\rm
v}=18.7\pm0.3$, in good agreement to my finding of $m_{\rm v}=19.07\pm0.52$,
which is represented by a S\'ersic profile.  Both the large uncertainty and
the large boxiness, $c=0.67\pm0.02$, are caused by the SBF.

\subsection {S\'ersic Bulge}

To test how different assumptions about the galaxy bulge can affect the
decomposition of P1 and P2, I redo the above fit by replacing the Nuker bulge
component with a S\'ersic model.  Figure 6 shows the surface brightness
profile and shape parameter plots similar to Figure 5.  The most prominent
difference between the two is the extent of the bulge profile slope inner to a
radius of $\approx2\arcsec$.  In terms of $\chi^2_\nu$, this fit is degenerate
with that presented in Figure 5.

Another notable difference between this decomposition and the previous is the
strength of the spherical bulge component, which now has a brightness of
$m_{\rm v}=13.51$.  Although this component is somewhat fainter than the
previous fit ($m_{\rm v}=12.76$), it is still a significant component in the
bulge relative to P1 and P2.

In this realization, the light from the two components P1 and P2 adds up to
$m_{\rm v} = 13.06$, which is nearly identical to the nominal fit with a Nuker
bulge above.  P1 has a magnitude $m_{\rm v} = 13.66$, again, nearly identical
to before.  This decomposition illustrates the main point of these different
exercises -- that throughout various trials, although the sub-components of P1
or P2, and even the bulge contribution, may not be unique, their summed flux
values and shapes are insensitive to model assumptions and initial parameter
values.  In the discussions below, I formally adopt the parameters from the
Nuker bulge decomposition because it gives a slightly better fit
statistically, and is a more general fit to the bulge, whereas the S\'ersic
profile has, by definition, a flat core.

\subsection {The Spherical Bulge Component}

In the previous two sub-sections, I preluded the presence of an underlying
spherical component embedded inside a much larger bulge, and is distinct from
the double nucleus.  With the bulge accurately removed, this extra component
has an axis ratio $q=0.97\pm0.02$, an effective radius of 12 pc (3\farcs2),
and close to an exponential disk profile.  In comparison, the bulge has an
axis ratio of $q=0.81\pm0.01$.  The spherical component half light radius is
roughly 5 times the separation of the double nucleus, moreover, contains 16\%
of the flux within 2$\arcsec$ (Table 1).  It is essential for a good fit, both
qualitatively and quantitatively, as shown throughout various trials above.

The spherical bulge component is not likely to be caused by dust in the bulge.
Figure 7 shows a bulge subtracted residual map of the entire WFPC2 mosaic in
the {\it V}-band.  Superimposed on the residual map are contours from the
image prior to subtraction.  The compass arrows show the major axis
orientation of the large-scale galactic disk and bulge.  Apparently, the bulge
isophotes are elliptical well into the central few arcseconds, where there is
little sign of dust.  In fact, Figure 8 shows a $V-I$ color map of the WFPC2
FOV, revealing strong color differences in areas affected by dust.
Remarkably, within $10\arcsec$ of the center, the presence of dust actually
diminishes.  Rather than due to dust, the existence of the spherical component
most likely results from a small but significant departure from the Nuker
profile and shape of the bulge.

Quillen, Bower, \& Stritzinger (2000), using {\it HST}\ NICMOS images,
discover that many galaxies, especially luminous core-types with boxy
isophotes, have rounder isophotes at small radii.  Peng et al. (2002) also
show that a spherical component can be accurately extracted from the centers
of some other galaxies.  There are two theoretical models that predict the
decrease in ellipticity and boxiness.  One involves the mixing of stochastic
orbits by the presence and growth of the central SBH (Norman, May, \& van
Albada 1985; Gerhard \& Binney 1985; and Merritt \& Valuri 1996), which
steepens the central cusp.  The other involves a dissipationless merging of
binary black holes that scatter stars from the center (Milosavljevic \&
Merritt 2001), thereby flattening the nuclear cusp, with a break radius that
can extend well beyond the sphere of influence of the SBH.  In Milosavljevic
\& Merritt (2001) model, binary black holes of similar mass can eject stars
amounting to the combined mass of the black holes.  Consequently, these
randomized stellar orbits, after redistributing in phase space, could
conceivably have produced a spherical bulge component, whose mass would then
be roughly that of the coalesced black holes.  The discovery of the spherical
component, in addition to the shallowness of the bulge cusp, suggest that the
binary merger scenario is a more natural explanation for the bulge formation.
Furthermore, the mass of the spherical component inferred from its luminosity
of $m_{\rm v} = 12.76$ is $2.8\times10^7 M_\odot$ -- a mass surprisingly
similar to the SBH of $3\times10^7 M_\odot$.  It remains to be seen whether
this agreement is a mere coincidence, or a confirmation that the bulge had
been scoured by merging binary black holes.

If the spherical bulge component resulted from a binary black hole merger
scenario, then the eccentric disk was formed more recently than the bulge.
The mixing of stellar orbits after a binary SBH merger would have erased any
organized sub-structure that previously resided at the nucleus.  This
hypothesis is consistent with the finding of SBV98, who discover that the
nuclear stars are significantly younger than stars in the bulge.

\section {LARGE SCALE BULGE PROPERTIES}

\subsection {Bulge Parameters and Correlations}

KB99 show a 1-D profile of the bulge, assembled piece-wise from several
studies, out to 6300\arcsec.  Their plot shows that the bulge has a gentle
curvature which has no meaningful break radius.  On the other hand, in my
representations of Figures 5 and 6, the bulge appears to have a triple power
law with a distinct break in the bulge profile near 2\arcsec\ which has not
been seen before because of the complications caused by the double nucleus
below $r=5\arcsec$.  From \S~4.3, I obtain 2-D Nuker parameters that are very
similar to those obtained by KB99, which is seen as a dashed line in Figure 5,
and which indeed does have a large break radius.  However, including the
spherical bulge reveals there to be a significant break at $r\approx2\arcsec$.
Whereas before, inferring from the Nuker fit alone, M31 bulge falls outside of
the correlation between $r_b$ with galaxy luminosity and $\mu_b$ found by
Faber et al. (1997), my bulge decomposition shows that the correlation now
holds very well.  My new estimates on the luminosity density ($j$), slope
($\gamma$), and the corresponding mass density ($\rho$), and slope ($\psi$)
values are listed in Table 2.  They, too, now fall within correlations with
galaxy luminosity shown in Faber et al.  (1997).

\subsection {Spiral Dust Structure and Color Gradient}

The decomposition of the mosaic $V$-band image (Figure 7) shows a beautiful
dust lane that runs from the lower-left of the image towards, but stopping
near, the center of the nucleus, and reappearing on the other side.  This
feature has previously been noted in ground based studies of Wirth, Smarr, \&
Bruno (1985) and SBV98.  In SBV98, the spiral dust arms extend out to roughly
30\arcsec\ on both sides of the nucleus.

The color map of Figure 8 reveals a significant gradient between the nucleus
and the bulge further out, toward the South.  It also highlights the winding
dust lanes towards the nucleus.  However, it is curious that despite the
extensive structure, the dust arms appear to stop within
$10\arcsec$-$15\arcsec$ of the center.  This will be discussed further in
\S~7.  The color difference between the nucleus and its immediate surrounding
noted by L98 and BA01 is evident.

\section {Formation of Eccentric Disk by Natural $m=1$ Mode vs. Disruption}

The mass of P1+P2 is instrumental for deciding between competing models that
form the double nucleus, in particular between the natural $m=1$ models (BA01
and SS01), and the star cluster disruption model of BE00.  All three produce
an eccentric disk as proposed by Tremaine, and broadly satisfy key dynamical
and morphological constraints.  In BE00 model, the key requirement is that the
globular cluster gets completely disrupted and scattered into a thick disk by
the SBH.  A massive cluster with strong self-gravity must approach the black
hole at a low impact parameter, which translates into a disk with high
ellipticity and a short life time.  To satisfy the morphological and dynamical
constraints, the cluster size is limited to $\lesssim 3\times10^6M_\odot$.  On
the other hand, the BA01 and SS01 models can be considerably more massive, in
the range 0.7-$2.1\times10^7M_\odot$.

My nominal integrated brightness for P1+P2 is $m_{\rm v} = 13.12\pm0.06$ which
translates to a mass $2.1\times10^7 M_\odot$.  In comparison to globular
clusters in M31 (Barmby, Huchra, \& Brodie 2001), P1 alone ($m_{\rm v} =
13.65$, $M=1.2\times10^7M_\odot$) would at least be in the 90 percentile of
the most luminous globular clusters.  Table 1 lists two sets of masses for
apertures $r=1\arcsec$ and $r=2\arcsec$\ centered on the UV peak, for all the
components.  I find that within $r=1\arcsec$ of the UV peak, the masses of the
components are $M_{bulge} = 8.3\times 10^6M_\odot$, $M_{P1} = 7.4\times 10^6
M_\odot$, and $M_{P2} = 5.0\times 10^6 M_\odot$.  The sum P1+P2 is comparable
to the black hole mass of $3\times10^7 M_\odot$.  For convenience, I also
provide masses within $2\arcsec$ aperture of the UV peak in Table 1.

The total mass of P1+P2 determined here is about an order of magnitude too
high for the globular cluster disruption scenario to work.  I note that P1
alone has sufficient mass to rule out this scenario, if P2 is at least in part
considered as the bulge.  The eccentric disk masses determined by L93 and K95
can rule out the globular cluster disruption scenario, as well.  BA01 also
conclude the disk is massive, $1.7\times10^7 M_\odot$, based on modeling the
bulge with multiple Gaussians.  My mass determination for the eccentric disk,
being higher than the first two studies, provides additional buffer to rule
out reasonable uncertainties in the M/L ratio by factors of a few.  On the
other hand, my P1+P2 flux is somewhat fainter (by 0.57 mag) than the
measurement of $m_{\rm v} = 12.55$ by KB99.  All estimates of the disk mass
are that it is massive, while Section 4 has shown that the uncertainty due to
the bulge decomposition amounts to about 6\%.  Therefore, the formation model
favored is that of BA01 or SS01.  In their models, an initial circular disk of
stars was formed in the bulge, but became more eccentric after a brief
encounter with a giant molecular gas cloud or a globular cluster, which need
not be disrupted.

The discovery of a spherical bulge component confirms that the bulge potential
is spherical around the double nucleus, which might be important for
sustaining a thin eccentric disk.  In a triaxial potential, an uniformly
precessing eccentric disk, held together by self-gravity, is subject to tidal
torque exerted by the bulge, which can cause misalignments in the orbital
configuration, or cause the orbits to diffuse into the bulge through phase
mixing.

\section {DISCUSSION and CONCLUSION}

I decompose M31 with GALFIT to accurately extract the double nucleus.  The
large scale bulge is made up of two components: a small, spherical structure,
embedded at the center of a large, and moderately elliptical component.  I
then use the decomposition to study the structural parameters of the bulge,
finding that the break radius, $r_b$, correlates with other galaxy parameters
that are found in Faber et al. (1997) for a large sample of early-type
galaxies.  The spherical component extracted has a mass of $2.8\times10^7
M_\odot$, which is surprisingly similar to the mass of M31's supermassive
black hole.  It remains to be seen whether this is the result of binary black
hole mergers, as predicted by Milosavljevic \& Merritt (2001) N-body
simulations, or a mere coincidence.

The inferred mass of P1 and P2 combined is comparable to the black hole mass,
and the bulge mass contained in a small region out to the radius of $1\arcsec$
from UV peak, with relative masses $M_\bullet:M_{bulge}:P1:P2 =
4.3:1.2:1:0.7$.  The inferred mass of P1+P2 is insensitive to the bulge
parameters, as others (e.g. BA01) had also found.  Coupled with the large
impact parameter suggested by the disk, it seems a cluster that resulted in
the eccentric disk of P1+P2 ($M\approx 2.1\times10^7 M_\odot$) could not have
been disrupted enough by the black hole ($M_\bullet\approx 3\times10^7
M_\odot$) to form an eccentric disk around it.  In such a scenario, the
progenitor of P1+P2 would have to be still larger than my mass estimate.

The large mass of the disk is consistent with a scenario in which a stellar
disk was formed in the nucleus, then became more eccentric after an $m=1$
perturbation, by, for example, passing giant molecular clouds or globular
clusters (BA01).  Another possibility is that an approaching globular cluster
of mass $\lesssim 10^6 M_\odot$ might have come within, and was shredded by,
the SBH, then was integrated into the pre-existing disk (SS01).  Despite how
well the current N-body simulations reproduce the kinematics and morphologies,
given that all the components within $2\arcsec$ of the bulge are comparable in
mass, a perturbation that would excite the $m=1$ mode in the disk might also
perturb the bulge.  Thus it is unclear whether their mutual tidal interactions
need to be further considered in N-body simulations, or might a smooth bulge
potential suffice.

The N-body simulations, coupled with a hypothetical shroud of dust at the
nucleus, might explain both the eccentric disk geometry as well as the slight
color difference between the bulge and the nucleus.  However, it can not be
the whole story:  KB99 show that P1 and P2 have metal line strengths stronger
than any globular cluster, hence they are unlikely to be the digested remnant
of a globular cluster or an elliptical galaxy.  A more complete picture also
needs to account for the differing bulge and nuclear stellar ages inferred by
SBV98.  One viable scenario as suggested by L98 is that the disk formation was
a separate event that occurred well after the formation of the bulge.  This
may be the case because the bulge formation event would have disrupted any
pre-existing sub-structure at the center.  To grow a disk to
$2.1\times10^7M_\odot$, one scenario is by the disruption of several globular
clusters.  However, this possibility is remote because angular momentum
conservation and scattering by the SBH would tend to produce a more spheroidal
geometry.  A more likely explanation, in private communication with L.~C. Ho,
is that a significant amount of gas and dust roughly $2.1\times10^7 M_\odot$
had accreted into the center.  Indeed, extended dust structures in Figures 7
and 8, as well as SBV98, strongly suggest it was a possibility.  Through
gravitational settling, a circular gas+dust disk enriched with reprocessed
material could have grown steadily, out of which young stars would then form.
This might explain the metallicity enhancement and the younger stellar
population in the nucleus compared to the bulge.  The disk might have
subsequently experienced a kick from a passing GMC or a globular cluster that
increased the eccentricity.

However, it is unclear at the moment how long ago or how such a massive
$2.1\times 10^7 M_\odot$ disk could have settled into the center.  Assuming
that the accretion occurred in the form of gas, the rate is limited by the
near absence of non-thermal AGN activity at the M31 nucleus.  Perhaps a
significant amount of wind from star formation near the center might quench
the AGN activity by blowing fuel away from the central engine (C.~D. Impey,
private communication).  The accretion + wind scenario is attractive for
explaining the possible presence of a UV star cluster at the center, and for
the absence of dust lanes in the immediate vicinity of the double nucleus,
despite there being a large and extended dust structure mere $20\arcsec$ away
(Figs. 7 and 8).

\bigskip

{\noindent \bf Acknowledgments.} I thank Peter Strittmatter for financial
support.  I also thank Luis Ho and Chris Impey for comments and reading the
original draft; Alice Quillen, and Dennis Zaritsky, and Roelof de Jong, for
general discussions; Daniel Eisenstein for assistance in the error
propagation; and Tod Lauer for providing the deconvolved images of M31, and
for discussions about the SBF and fitting of the double nucleus.  I also thank
the referee for helpful and insightful comments.  This research has made use
of the NASA/IPAC Extragalactic Database (NED) which is operated by the Jet
Propulsion Laboratory, California Institute of Technology, under contract with
the National Aeronautics and Space Administration.


\begin{deluxetable}{llrrrrrrrrrrrrrl}
\rotate
\tablewidth{0pt}
\tablecaption {Two Dimensional Image Fitting Parameters}
\tablehead{ Trial & Func. & $\Delta\alpha$ & $\Delta\delta$ & $f/f_{tot}$ & $f/f_{tot}$ &
m$_{\rm v}, \mu_{\rm v}$ & $r_{\hbox{b,e}}$ & $\alpha, n$ & $\beta$ & $\gamma$ & $a/b$ & PA & $c$ & $\chi^2_\nu$ & Comments \\
m$_{\rm v}(r\le1\arcsec\ \&\ 2\arcsec)$  &     & ($\arcsec$) & ($\arcsec$) & r$\le1\arcsec$ & r$\le2\arcsec$ & (mag) & ($\arcsec$) & & & & & (deg) & & \\
Mass$(r\le1\arcsec\ \&\ 2\arcsec)$ \\
(1)    & (2) & (3) & (4) & (5) & (6) & (7) & (8) & (9) & (10) & (11) & (12) & (13) & (14) & (15) & (16) }
\startdata
Solution 1:               & S\'ersic &  $\equiv0.$ & $\equiv 0.$ & 0.004 & 0.002  & 19.07 &   0.08 & 0.42 &      &      & 0.63 &   70.4  &  0.67 & 1.30 & UV peak  \\
Nuker bulge               & S\'ersic &     $-0.24$ &    $-0.15$  &  0.24 &  0.15  & 14.17 &   0.87 & 0.98 &      &      & 0.66 &   54.3  &  0.05 &      & P2  \\
                          & S\'ersic &        0.31 &       0.42  &  0.26 &  0.16  & 13.88 &   1.00 & 1.31 &      &      & 0.62 &   64.8  & $-0.24$&     &  P1a \\
13.10, 12.20 mag          & S\'ersic &        0.49 &       0.37  &  0.08 &  0.05  & 15.46 &   0.37 & 0.69 &      &      & 0.92 & $-46.8$ &  0.57 &      & P1b \\
                          & S\'ersic &        0.13 &    $-0.06$  &  0.11 &  0.16  & 12.76 &   3.21 & 0.83 &      &      & 0.97 &   56.0  &  0.14 &      & Spherical bulge \\
$2.07\times10^7 M_\odot$, & Nuker    &     $-0.06$ &    $-0.15$  &  0.29 &  0.45  & 17.76 &  66.48 & 1.10 & 1.99 & 0.17 & 0.81 &   50.1  &  0.05 &      & Bulge\\
$4.75\times10^7 M_\odot$  & Offset   &             &             &  0.01 &  0.01  & 20.4  &        &      &      &      &      &         &       &      & mag/arcsec$^2$ \\
\tableline
Solution 2:               & S\'ersic & $\equiv 0.$ & $\equiv 0.$ & 0.005 & 0.002  & 18.86 &   0.09 & 0.41 &      &      & 0.67 &   60.7  &  1.99 & 1.30  & UV peak \\ 
S\'ersic bulge            & S\'ersic &     $-0.22$ &     $-0.14$ &  0.27 &  0.18  & 13.98 &   0.90 & 1.00 &      &      & 0.69 &   55.6  &  0.01 &       & P2 \\   
                          & S\'ersic &        0.33 &       0.43  &  0.28 &  0.20  & 13.83 &   0.90 & 1.20 &      &      & 0.66 &   64.9  & $-0.26$&      & P1a \\
13.10, 12.20 mag          & S\'ersic &        0.51 &       0.38  &  0.07 &  0.04  & 15.77 &   0.36 & 0.68 &      &      & 0.88 & $-44.7$ &  0.83 &       & P1b \\
                          & S\'ersic &        0.00 &       0.16  &  0.08 &  0.11  & 13.51 &   2.40 & 0.44 &      &      & 0.96 &   56.6  &  0.08 &       & Spherical bulge \\
$2.07\times10^7 M_\odot$, & S\'ersic &        0.10 &    $-0.17$  &  0.29 &  0.47  & 4.97  & 214.71 & 2.00 &      &      & 0.81 &   52.0  &  0.06 &       & Bulge \\
$4.75\times10^7 M_\odot$  & Offset   &             &             &  0.01 &  0.01  & 20.4  &        &      &      &      &      &         &       &       & mag/arcsec$^2$ \\
\tableline
Uncertainties             &          &        0.02 &       0.02  &       &        & 0.2 &   2\%  & 0.03 &  0.02  & 0.02 & 0.03 &      2  &  0.02 &       &\\
\tableline
\enddata 
\tablecomments { 
Col. (1): Fitting trials for Nuker or S\'ersic bulge type, apparent visual
	  magnitude within an aperture $1\arcsec$ and $2\arcsec$ radius
	  centered on the UV peak, and the corresponding inferred {\it
	  luminous} mass based on $M/L=5.7$.
Col. (2): Galaxy components used in the fit.  
Col. (3): RA offset.  
Col. (4): DEC offset.  
Col. (5): The fraction of component flux with respect to the total integrated
	  flux (Col. 1), all within a $r=1\arcsec$ aperture centered on the UV
	  peak.
Col. (6): Same as Col. 5, but with $r=2\arcsec$ aperture.
Col. (7): For Nuker, it is the surface brightness at the breaking radius.  For
	  the S\'ersic profile it is the total brightness.  For the {\it
	  offset} component (which represents an exponential disk) it is the
	  surface brightness magnitude.  The magnitudes are not corrected for
	  galactic extinction.
Col. (8): $r_{\hbox{b}}$ is the breaking radius for the Nuker power law,
	  $r_{\hbox{e}}$ is the effective radius of the S\'ersic law.
	  Both have units in arcseconds.
Col. (9): For Nuker, $\alpha$ parameterizes the sharpness of the break.  For
          S\'ersic, $n$ is the S\'ersic exponent 1/$n$.
Col. (10): Nuker asymptotic outer power law slope.  
Col. (11): Nuker asymptotic inner power law slope.  
Col. (12): Axis ratio.  
Col. (13): Position Angle.  
Col. (14): Diskiness (negative)/boxiness (positive) parameter.  
Col. (15): Reduced $\chi^2$ of the fit.
Col. (16): Comments:  P1a and P1b add to form a single component P1.
The last row shows the ``representative'' uncertainties for the ensemble
of profiles.  Individual uncertainties are quoted in the text where
appropriate.
}
\end{deluxetable}

\begin{deluxetable}{lcccccccc}
\tablewidth{0pt}
\tablecaption {Nuclear Luminosity/Mass Density Parameters}
\tablehead{Bulge Type & 
$\left< \gamma \right>$ & $\left<\gamma\right>$ & 
log$\left<j\right>$  & log$\left<j\right>$ & 
log$\left<\rho\right>$  & log$\left<\rho\right>$ & 
$\left<\psi\right>$     & $\left<\psi\right>$ \\
 & $(r<0\farcs1)$ &  $(r<10{\rm pc})$ & $(r<0\farcs1)$ &  $(r<10{\rm pc})$ &
$(r<0\farcs1)$ &  $(r<10{\rm pc})$ & $(r<0\farcs1)$ &  $(r<10{\rm pc})$ \\
(1) & (2) & (3) & (4) & (5) & (6) & (7) & (8) & (9)\\}
\startdata
Nuker    & 0.14 & 0.15 & 3.97 & 2.65 & 4.73 & 3.41 & 1.16 & 1.12 \\
S\'ersic & 0.01 & 0.03 & 3.33 & 2.59 & 4.09 & 3.35 & 0.51 & 0.51 \\
\tableline
\enddata
\tablecomments { 
Col. (1):  The bulge is defined as the sum of two components described in the
         text: the spherical + Nuker, or spherical + S\'ersic.
Col. (2 and 3): The average logarithmic power law slope of the bulge surface
		brightness ($\left<d\mbox{log}I/d\mbox{log }r\right>$) within
		$r < 0\farcs1$ and 10pc.
Col. (4 and 5): The logarithmic average luminosity density $j$ 
		integrated within $r < 0\farcs1$ and 10pc, in units of
		[$L_\odot/{\rm pc}^3$].
Col. (6 and 7): The logarithmic mass density $\rho$ inferred from
                Col. 4 and 5, in units of [$M_\odot/{\rm pc}^3$], based
                on $M/L=5.7$.
Col. (8 and 9): The average mass density power law slope $\psi$
		($\left<d\mbox{log}\rho/d\mbox{log }r\right>$), analogous to
		$\left< \gamma \right>$.
}
\end{deluxetable}


\begin{figure} 
    \plotone {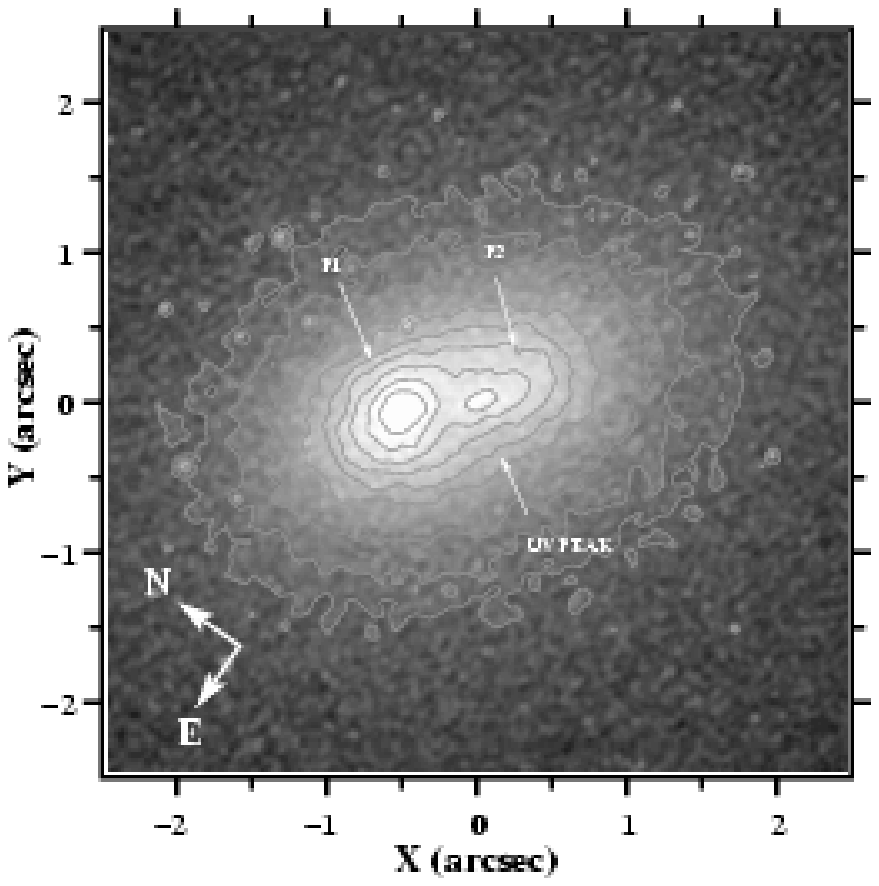} 

    \figcaption[cpeng-m31.fig1.eps] {A deconvolved (L98) grey scale image of
    the M31 double nucleus in the $V (F555W)$ band.  The intensity stretch is
    logarithmic, and the contour interval is 0.2 mag.}

\end{figure}

\begin{figure} 
    \plotone {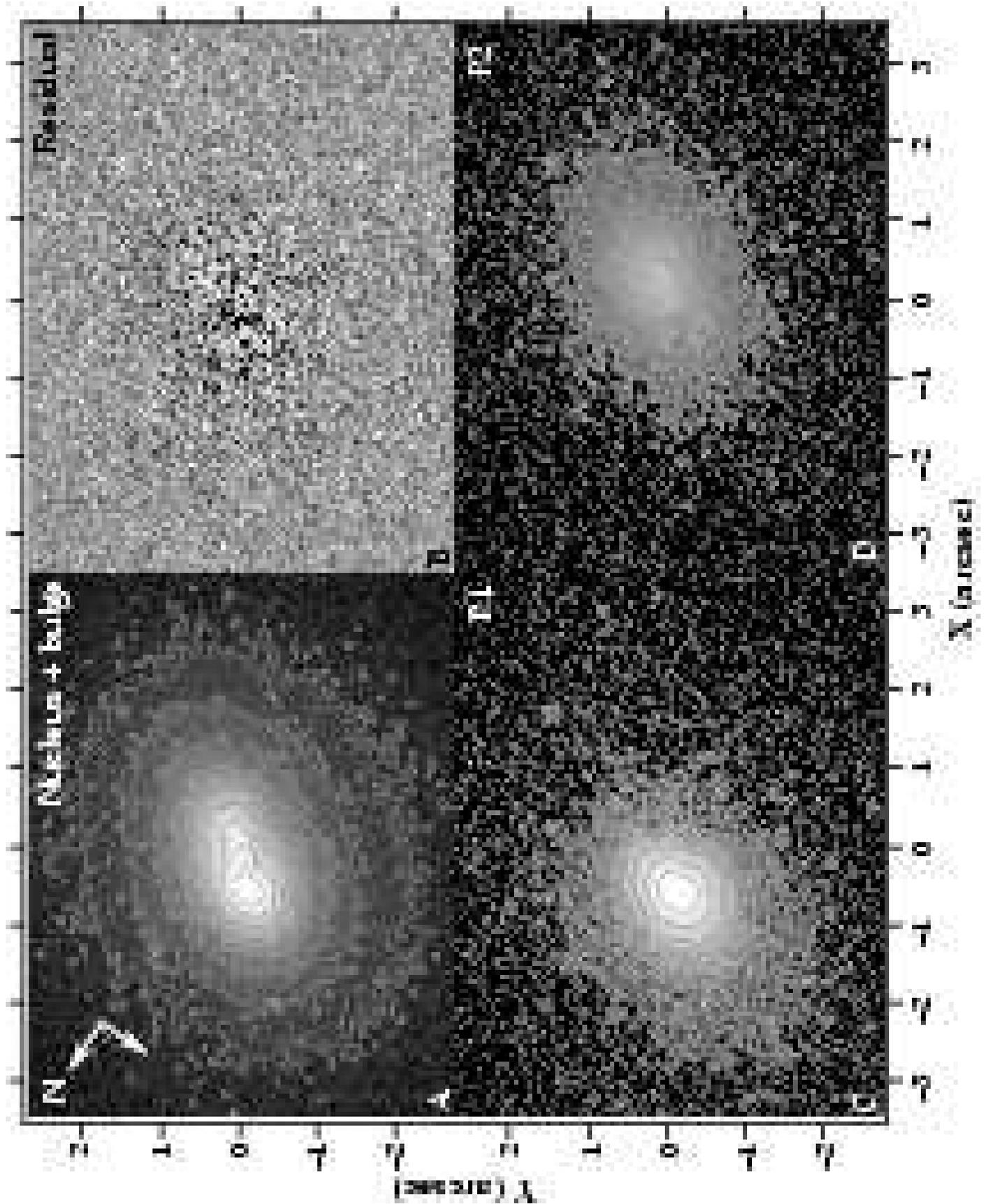} 

    \figcaption[cpeng-m31.fig2.eps] {Decomposition of M 31.(a) bulge + nucleus.
    (b) residuals from 2-D GALFIT are shown in positive grey scale.  (c) P1
    component.  (d) P2 component. The contour intervals are arbitrary.}

\end{figure}

\begin{figure}
    \plotone {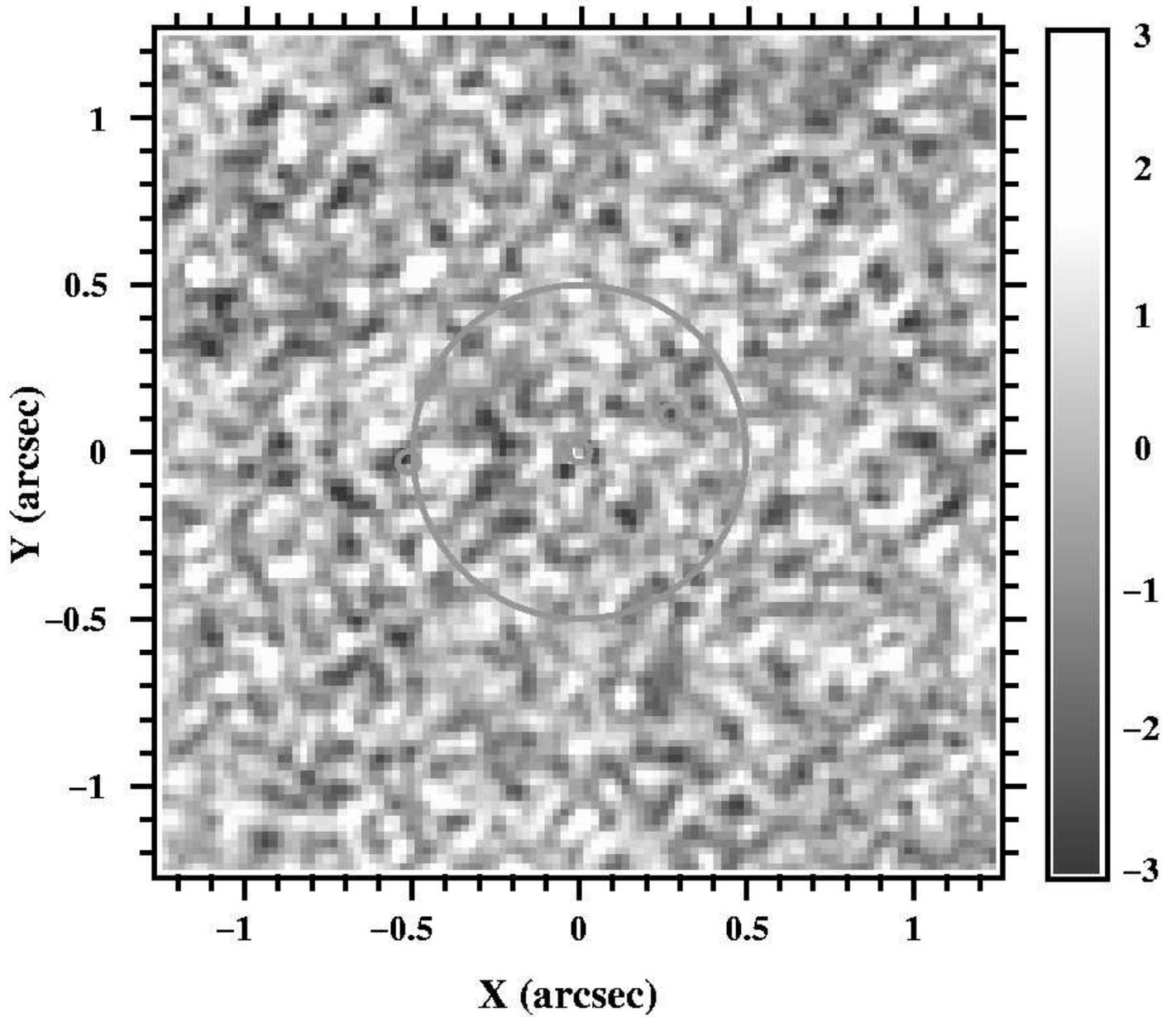}

     \figcaption[cpeng-m31.fig4.eps] {Image of the fit residuals divided by
     the local RMS, with intensity scale shown to the right.  The calculation
     of the RMS image is discussed in \S~S3.2.  The large circle has a radius
     of $0\farcs5$ centered on the UV peak (small central circle).  The other
     two circles at ($-0\farcs51$, $-0\farcs02$) and ($0\farcs28$, $0\farcs1$)
     represent the position of P1 (left) and P2 (right), respectively.  The
     image orientation is the same as Figs. 1 and 2.}

\end{figure}

\begin{figure} 
    \plotone {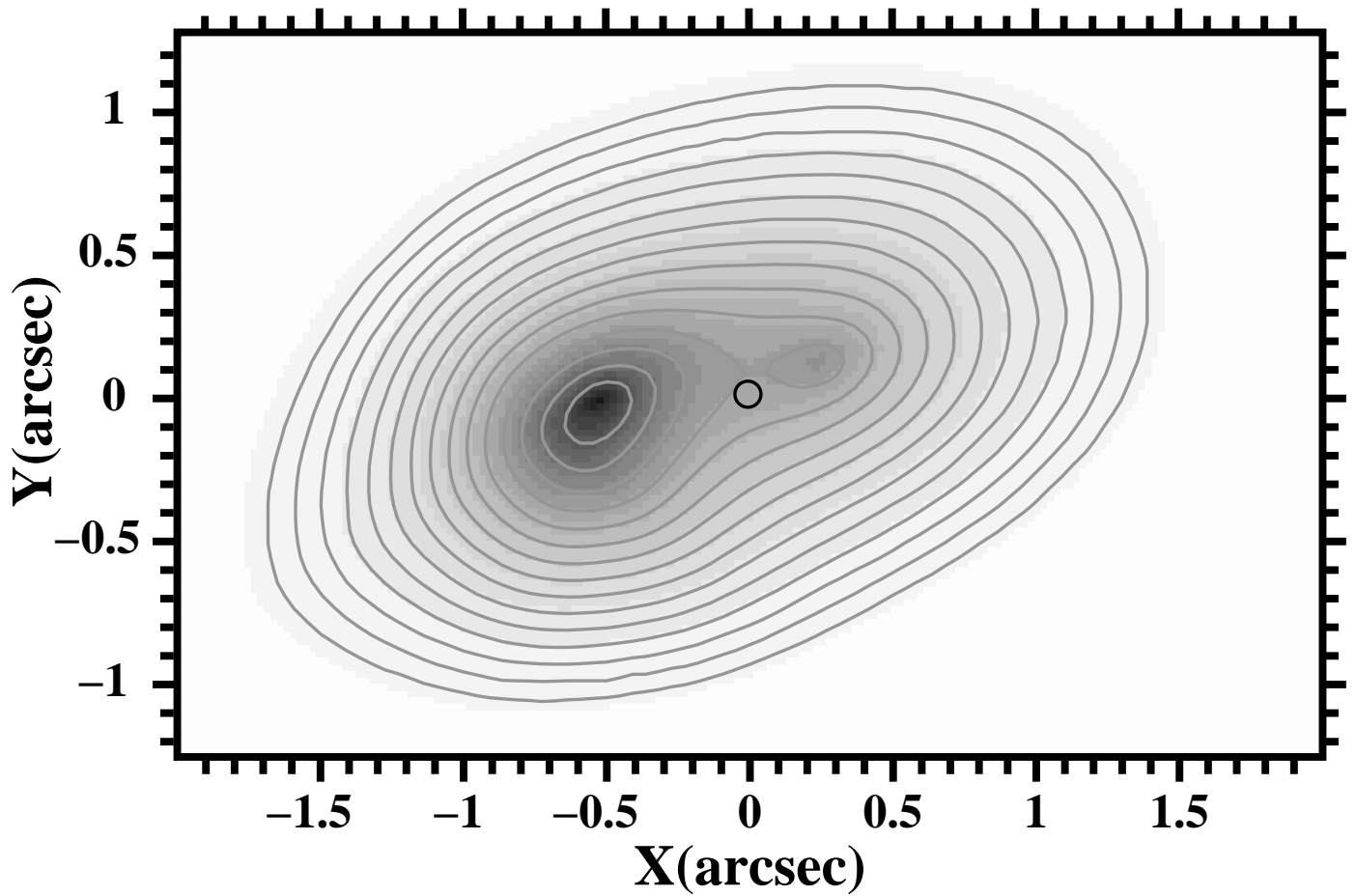} 

    \figcaption[cpeng-m31.fig3.eps] {Model of the eccentric disk, P1 and P2. 
    The position of the UV peak is marked by a circle at the center.  The
    contour interval is 0.1 magnitude.}

\end{figure}

\begin{figure}
    \plotone {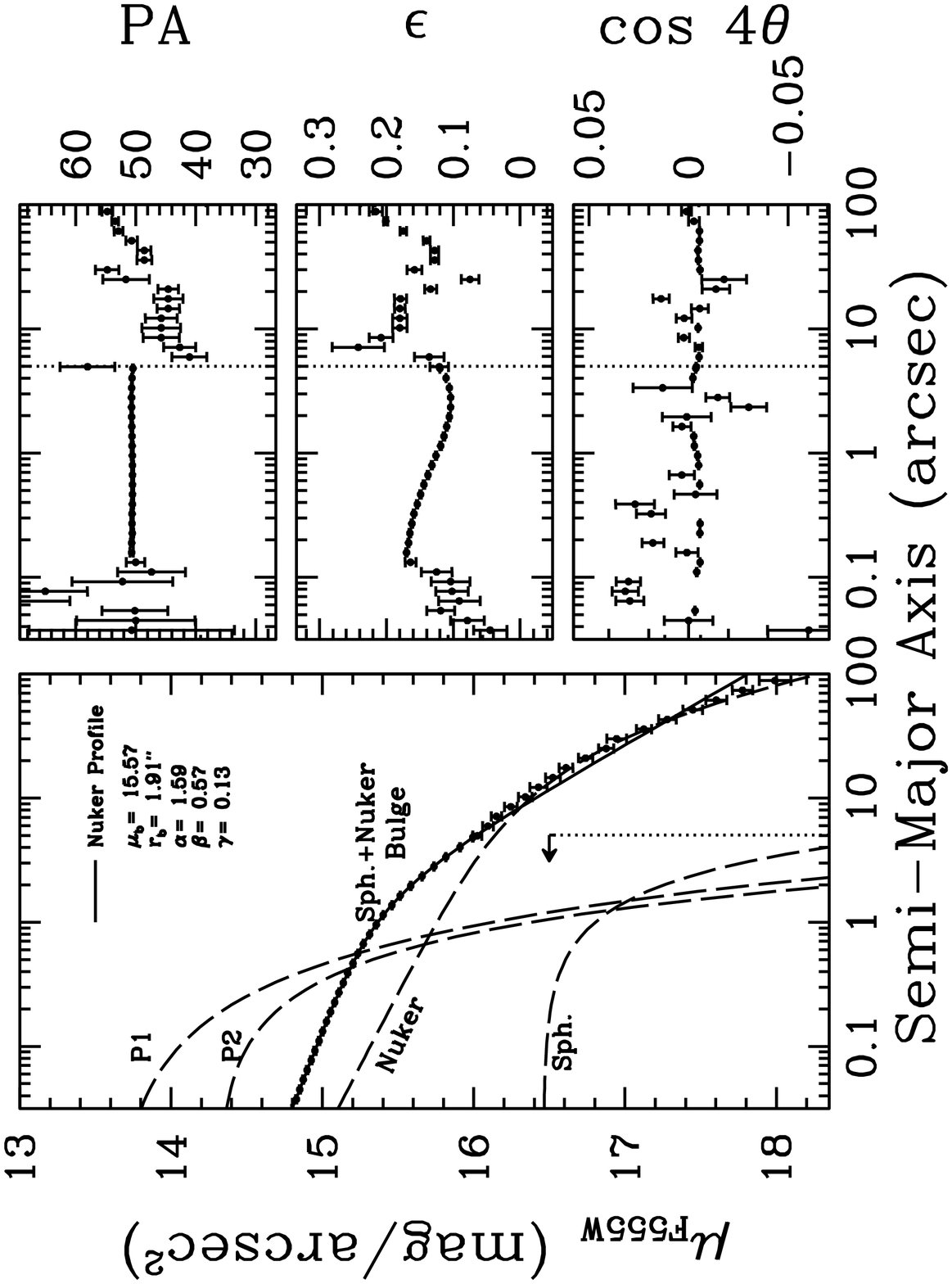} 

    \figcaption[cpeng-m31.fig5.eps] {Isophote fits to the WFPC2 mosaic image.
    The surface brightness profile panel shows the ``intrinsic'' bulge profile
    as data points.  Inner to $5\arcsec$, marked by vertical dotted lines, the
    double nucleus begins to dominate, so we replace that region by a sum of
    the fitted Nuker + spherical bulge components (see text).  The profile of
    the individual components used in the fit are shown in dashed lines.  The
    solid line running through the data points is a 1-D Nuker function with
    parameters embedded in the figure.  In the right-hand panels, PA is the
    position angle, $\epsilon$ is the ellipticity, and $\mbox{cos} 4\theta$ is
    the diskiness ($>0$) and boxiness parameter ($<0$).}

\end{figure}

\begin{figure}
    \plotone {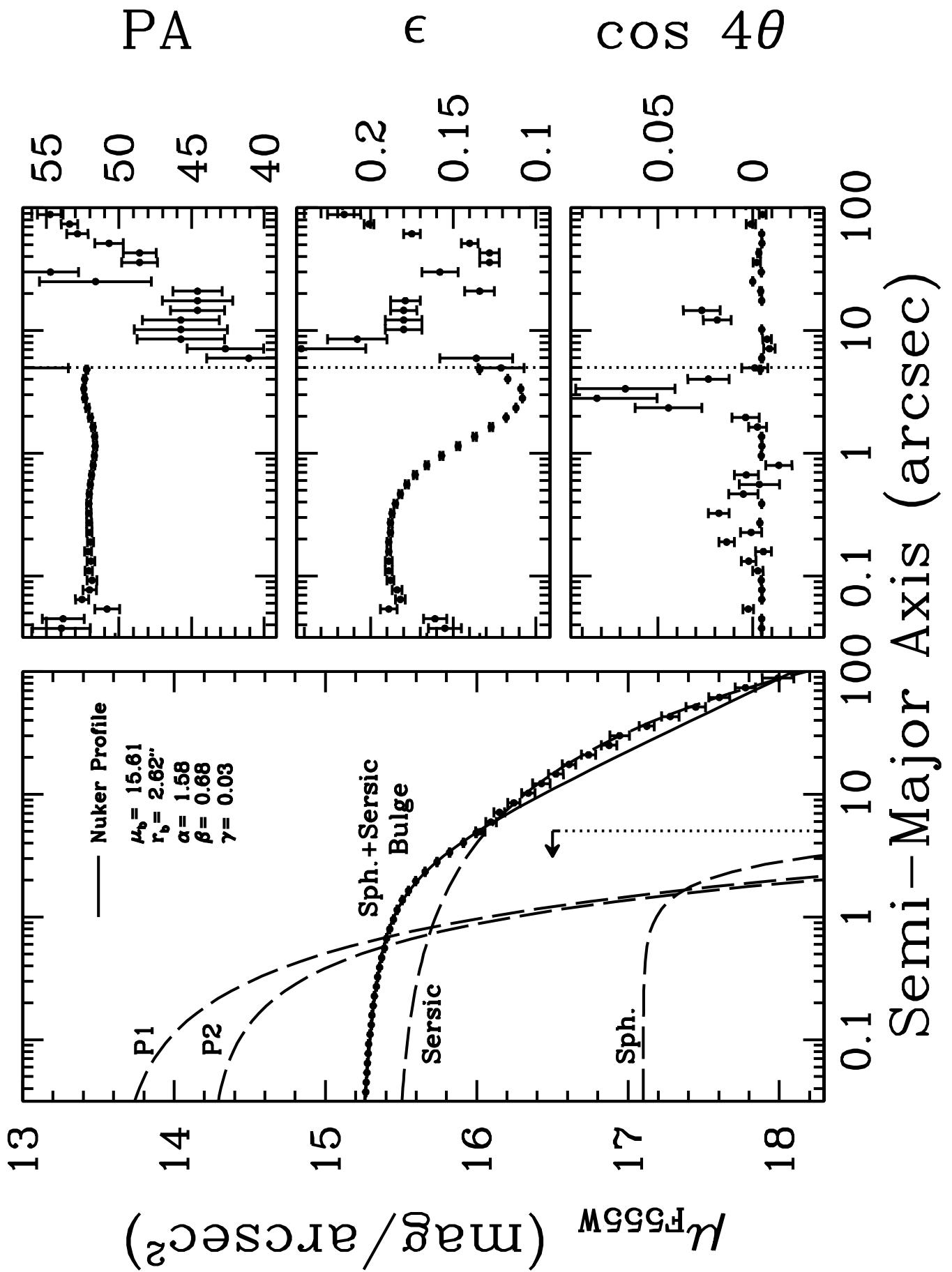} 

    \figcaption[cpeng-m31.fig6.eps] {Isophote fits to the WFPC2 mosaic image.
    The surface brightness profile panel shows the ``intrinsic'' bulge profile
    as data points.  Inner to $5\arcsec$, marked by vertical dotted lines, the
    double nucleus begins to dominate, so we replace that region by a sum of
    the fitted S\'ersic + spherical bulge components (see text).  The profile
    of the individual components used in the fit are shown in dashed lines.
    The solid line running through the data points is a 1-D Nuker function
    with parameters embedded in the figure.  In the right-hand panels, PA is
    the position angle, $\epsilon$ is the ellipticity, and $\mbox{cos}
    4\theta$ is the diskiness ($>0$) and boxiness parameter ($<0$).}

\end{figure}

\begin{figure} 
    \plotone {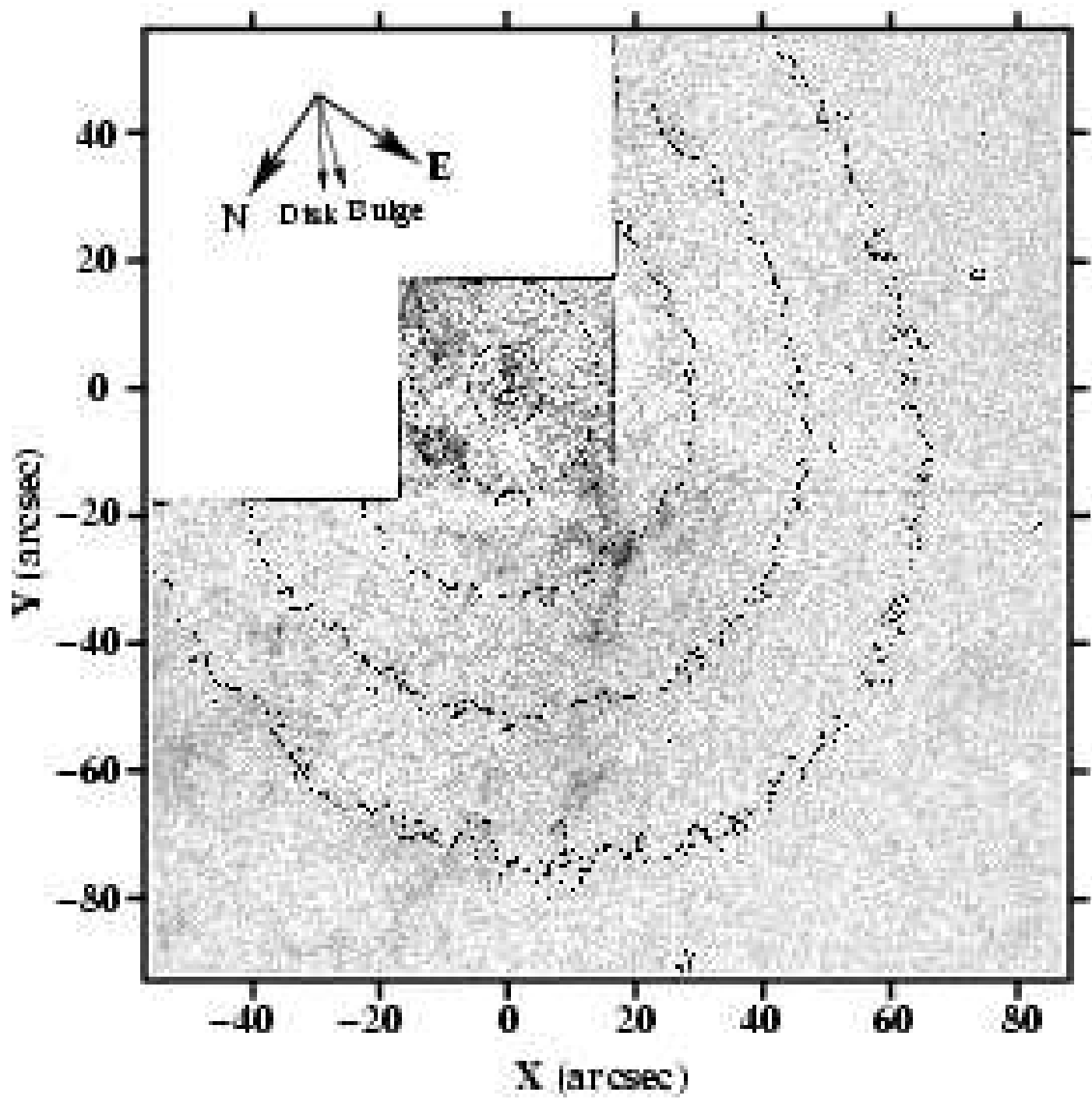} 

    \figcaption[cpeng-m31.fig7.eps] {A residual image with the bulge and
    nucleus subtracted, showing a dust lane that runs clear through the
    center.  The isophotes superposed are from the same image before
    subtraction, and have logarithmic spacing.  The compass shows the
    orientation of the image with respect to the sky, as well as the major
    axis orientation of the M31 galactic disk and large scale bulge.}

\end{figure}

\begin{figure}
    \plotone {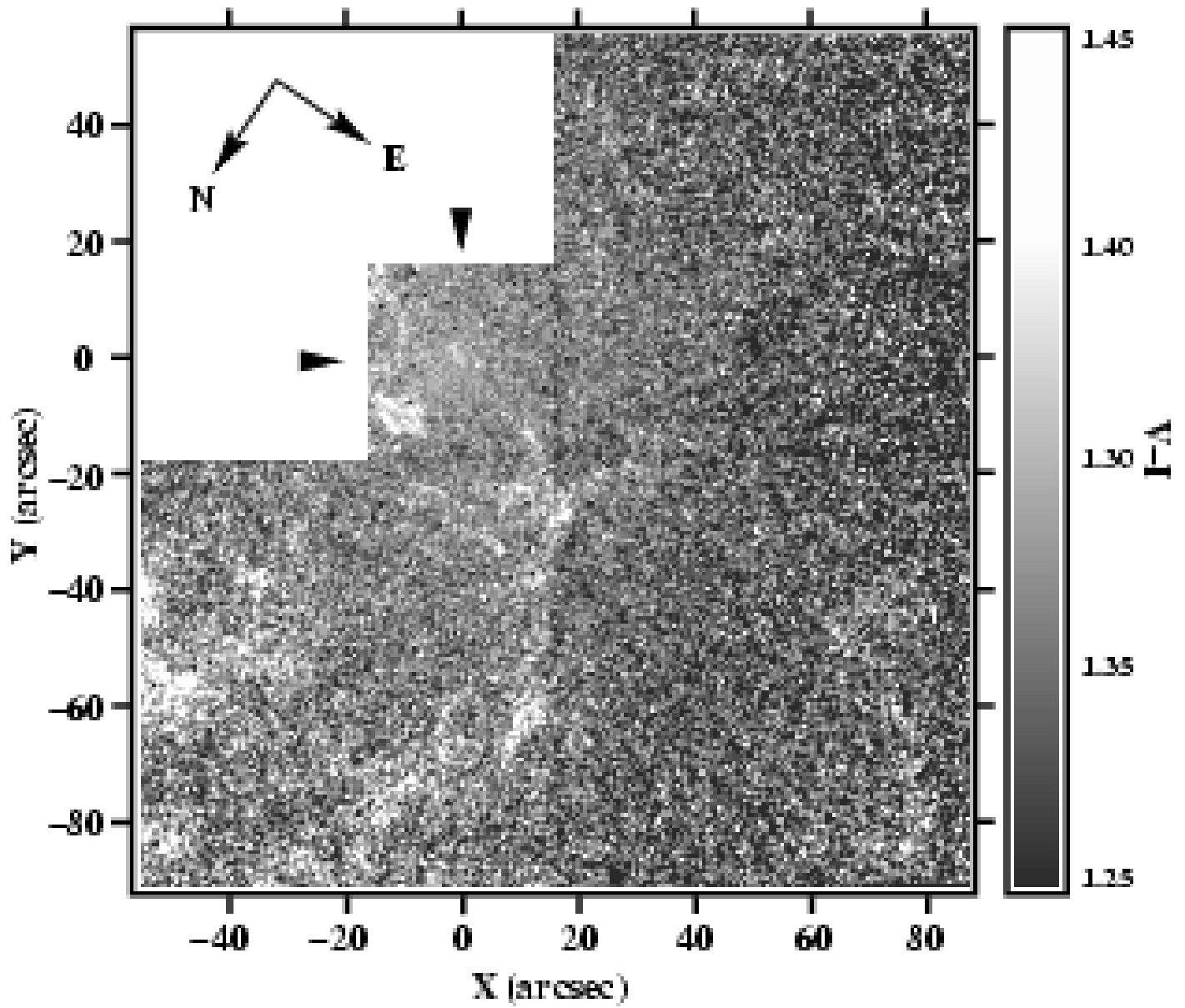} 

    \figcaption[cpeng-m31.fig8.eps] {A ($V-I$) color map showing color
    gradient of the bulge.  The arrowheads mark the location of the double
    nucleus.  The image has been smoothed over by a Gaussian kernel of
    $\sigma=2$ pixels.}

\end{figure}

\end {document}